
\magnification \magstep1
\raggedbottom
\openup 2\jot
\voffset6truemm
\headline={\ifnum\pageno=1\hfill\else
{\it SELF-DUAL ACTION FOR FERMIONIC FIELDS AND GRAVITATION}
\hfill \fi}
\rightline {June 1994, DSF preprint 94/23}
\centerline {\bf SELF-DUAL ACTION FOR FERMIONIC FIELDS}
\centerline {\bf AND GRAVITATION}
\vskip 1cm
\centerline {\bf Hugo A. Morales-T\'ecotl$^{1,2}$ and
Giampiero Esposito$^{3,4}$}
\vskip 1cm
\centerline {\it ${ }^{1}$Universidad Aut\'onoma Metropolitana
Iztapalapa$^{\ast}\!\!,$}
\centerline {\it Departamento de F\'{\i}sica,
Apartado Postal 55-534,}
\centerline {\it 09340 M\'exico D.F., M\'exico;}
\centerline {\it ${}^2$SISSA/ISAS, Via Beirut 2-4, 34013 Trieste, Italy;}
\centerline {\it ${ }^{3}$Istituto Nazionale di Fisica Nucleare}
\centerline {\it Mostra d'Oltremare Padiglione 20, 80125 Napoli, Italy;}
\centerline {\it ${ }^{4}$Dipartimento di Scienze Fisiche}
\centerline {\it Mostra d'Oltremare Padiglione 19, 80125 Napoli, Italy.}
\vskip 1cm
\noindent
{\bf Summary. -}
This paper studies the self-dual Einstein-Dirac theory.
A generalization is obtained of the Jacobson-Smolin proof
of the equivalence between the self-dual
and Palatini purely gravitational actions.
Hence one proves equivalence of self-dual Einstein-Dirac
theory to the Einstein-Cartan-Sciama-Kibble-Dirac theory.
The Bianchi symmetry of the curvature,
core of the proof, now contains
a non-vanishing torsion.
Thus, in the self-dual framework,
the ``extra" terms entering the equations of motion
with respect to the standard Einstein-Dirac
field equations, are neatly associated with torsion.
\vskip 1cm
\leftline {$^{\ast}$ Present E-mail address: hugo{@}xanum.uam.mx}
\vskip 1cm
\leftline {PACS numbers: 04.20.Cv}
\vskip 100cm
\leftline {\bf 1. - Introduction.}
\vskip 0.3cm
Over the last few years, many efforts have been produced in
the literature to provide a new mathematical framework for the
analysis of canonical gravity (e.g. [1-2] and references therein).
Thus, instead of the standard geometrodynamical variables, one
now deals with a new set of variables involving soldering
forms and connections. Remarkably, in terms of these geometrical
objects proposed by Ashtekar [1],
the constraints of general relativity take a polynomial
form, and this has motivated the introduction of yet new
(loop) variables to solve the quantum version
of the constraint equations [1].

So far, any testable property of gravitation
involves matter, and, in fact, matter could help to come to terms
with the issue of defining physical observables and time
in quantum gravity [3]. Hence,
coupling matter fields to gravitation becomes of primary interest.
Spin-${1\over 2}$ fields coupled to gravity
within the framework of Ashtekar's
variables were studied in [4] and
[5] (the latter included, besides, scalar and Yang-Mills fields).
Since the gravitational part of the
action using these variables is first-order, it is natural to
consider a gravitational connection
admitting torsion.
There are several kinds of matter which
can support a non-vanishing torsion
when coupled to gravity [5-6]. Attention is here focused
on the spin-${1\over 2}$ Dirac field minimally coupled to gravity.
A key question is whether the simplifying
Ashtekar form of the theory contains the whole information of the
standard well-known theories, at the classical level.
In the pure-gravity case, Jacobson and Smolin [2]
showed that the Palatini action
$$
S_{P}[e,\omega]= \int d^4\!x\;\; e \; e^{a\hat{a}} e^{b\hat{b}}
R_{ab\hat{a}\hat{b}}(\omega)\;,
\eqno (1.1)
$$
with $e^{-1}$ the determinant of the tetrad
$e^{a\hat{a}}$, $a,b$ world indices and $\hat{a},\hat{b}$ frame
indices, can be recovered from the self-dual action
$$
S_{SD}[e, {^+}\omega] = \int d^4\!x\;\; e \; e^{a\hat{a}} e^{b\hat{b}}
R_{ab\hat{a}\hat{b}}({^+}\omega)\; ,
\eqno (1.2)
$$
with ${^+}\omega:= {1\over 2} (\omega-i\ast\omega)$ the
self-dual (on frame indices) part of $\omega$. This follows on
substituting the equation of motion $\delta S_{SD}/\delta{{^+}\omega}=0$
in (1.2) [2]. That is to say
$$
S_{SD}[e,{^+}\omega(e)]= {1\over 2} S_{P} [e,\omega(e)]
- {i\over 4} \int d^4\!x\;\; e \; e^{a\hat{a}}
\epsilon_{\hat{a}}^{\;\;\hat{c}\hat{d}\hat{e}}
R_{a\hat{c}\hat{d}\hat{e}}(\omega(e))\; .
\eqno (1.3)
$$
Thus, by virtue of the Bianchi symmetry of the curvature
for vanishing torsion,
$R_{a[\hat{c}\hat{d}\hat{e}]}=0$, the second term in (1.3) vanishes.

Our work aims to complement the results of [4-5], on
spin-${1\over 2}$ fields, in two respects:
\vskip 0.3cm
\noindent
(i) The equivalence (analogue of (1.3)) is {\it explicitly} given for
the ECSKD (Einstein-Cartan-Sciama-Kibble-Dirac) [6-7]
and the sd-ED (self-dual Einstein-Dirac) actions [4-5].
\vskip 0.3cm
\noindent
(ii) The origin of some extra terms in the equations
of motion derived from the self-dual action [5],
w.r.t. standard Einstein-Dirac ones [8], is traced back to the
non-vanishing torsion. A result considered previously, in
a different framework, in the literature [9].

Moreover, our proof of the equivalence (i) is
necessarily a generalization of that of Jacobson and Smolin [2]
since, for non-vanishing torsion, $R_{a[cde]}\ne 0$. Nevertheless,
the Bianchi symmetry of the curvature including
torsion already exists [10] and we make use of it below.

Sect. {\bf 2} establishes
the equivalence between the sd-ED and ECSKD actions.
Sect. {\bf 3} studies the field equations for both theories.
Concluding remarks are presented in sect. {\bf 4}.
\vskip 0.3cm
\leftline {\bf 2. - ECSKD theory and self-dual Einstein-Dirac theory.}
\vskip 0.3cm
In coupling fermionic fields to gravity the introduction of orthonormal
tetrads is natural because spinors are defined w.r.t. orthonormal frames
[10]. Furthermore, whenever tetrads are adopted,
connections also enter
the description of gravity. In building up an action from which
to obtain the equations of motion, one has the possibility of considering
tetrads and connections as independent fields or not. If they are not
one gets the Einstein-Dirac (second-order) action. Instead, by taking
them as independent, one gets the
ECSKD (first-order) action. The corresponding variational problems
differ on what should be fixed at the boundary. One finds it is
necessary to add boundary terms to get a well-posed problem
only in the second-order case [11].
In the first-order case, on the other hand, one is left
with an equation of motion associating a non-vanishing torsion to the
connection [6-7].

In a four-dimensional Lorentzian space-time, a massive Dirac
field is represented by the spinor fields $\Bigr(\kappa^{A},
\mu^{A}\Bigr)$, say, jointly with their complex conjugates,
hereafter denoted by overbars. Thus, using two-component
spinor notation, the corresponding action functional for
ECSKD theory is
$$ \eqalignno{
S_{ECSKD} &= \int_{M} d^{4}x \left\{ \sigma
\left[\sigma^{aMA'} \sigma^b_{~AA'} R_{abM}^{~~~~A}
[{ }^{+}\omega]
+ \sigma^{aAM'} \sigma^b_{~AA'}
\overline{R}_{abM'}^{~~~~A'}[{ }^{-}\omega]
\right]\right. \cr
&- \sqrt{2}\;\sigma\; \sigma^{a}_{~AA'} \left[
\overline{\kappa}^{A'} \left( \nabla_a \kappa^A \right)
- \left(\nabla_a \mu^A\right) \overline{\mu}^{A'} \right]\cr
&+ \sqrt{2}\;\sigma\; \sigma^{a}_{~AA'} \left[
\left(\nabla_a \overline{\kappa}^{A'} \right) \kappa^{A}
- \mu^A \left(\nabla_a \overline{\mu}^{A'} \right)
\right] \cr
& \left. - 2im \sigma \left[ \mu_A \kappa^A -
\overline{\kappa}^{A'} \overline{\mu}_{A'}\right] \,
\right\} ,
&(2.1)\cr}
$$
where $\sigma \equiv {\rm det}(\sigma_a^{\;\;AA'})$
and $\sigma_a^{\;\;AA'}$ is the soldering form
(i.e. the two-spinor version of the tetrad in curved
space-time).
Note that the connection, here splitted into self-dual and
anti-self-dual parts, develops a torsion
contribution supported by the fermionic fields,
since we use a first-order
formalism with connection and soldering forms taken to be independent
fields instead of adopting {\it a priori}
the relation between them [8].

By contrast, the authors of [5]
studied the coupling of fermions to gravity in terms
of an action containing the self-dual part of a connection only,
disregarding the anti-self-dual part.
They assumed, though, this connection to be torsion-free.
In this paper the
analysis is carried out by extending the connection to admit torsion.
This simplifies the proof of equivalence between sd-ED and ECSKD,
as shown below, and makes it easier to interpret the corresponding
equations of motion in section 3.
Let the self-dual action [5] be
$$ \eqalignno{
S_{SD} &= \int_{M} d^{4}x \left\{- \sigma \;
\sigma^{aMA'} \; \sigma^{bJ}_{\;\;A'} \; F_{abMJ}
\right. \cr
&- \sqrt{2} \; \sigma \; \sigma^{a}_{~AA'} \left[
\overline{\kappa}^{A'}
\left({\cal  D}_a \kappa^A \right)
-\left({\cal  D}_a \mu^A\right)
\overline{\mu}^{A'}\right]\cr
& \left. - im \sigma \left[ \mu_A \kappa^A -
\overline{\kappa}^{A'} \overline{\mu}_{A'}\right] \,
\right\} ,
&(2.2)\cr}
$$
where $F_{abMN}$ is the curvature of the connection $\cal D$ defined
at this stage to act only on unprimed spinor indices. Clearly,
(2.2) is obtained from (2.1)
by taking only the contribution of the {\it self-dual}
piece of the connection, ${ }^{+}\omega$
and half of the mass term. The name {\it self-dual} hence accounts
for this. Thus, (2.2) is manifestly
not real. Nevertheless, it will be shown below it reproduces
(2.1) modulo the equations of motion for $\cal D$
and via the Bianchi
symmetry of the curvature for a ``metric-compatible" connection having
torsion, so that no spurious equations of motion are picked up.

The goal here is to determine $\cal D$ dynamically.
The variation of (2.2) with respect to
${\cal D}$ can be
carried out by introducing the auxiliary forms $Q_{a~N}^{~M}$
and $P_{\;\;ab}^{c}$ so as to define
${\cal D}$ with respect to
$\nabla$, the connection compatible with the soldering form,
i.e. such that
$
\nabla_a  \sigma_b^{AA'} = 0,
$
and having associated a non-vanishing torsion $T_{ab}^{~~c}$
$$
2 \nabla_{[a}\nabla_{b]} f \equiv T_{ab}^{~~c} \; \nabla_c f\, , \,
f\, {\rm a~ zero-form} .
\eqno (2.3)
$$
Namely,
$$
{\cal D}_a \lambda^A_b = \nabla_a \lambda^A_b
+ Q_{a~B}^{~A} \; \lambda_{b}^{B}
+ P_{\;\;ab}^{c} \; \lambda^{A}_{c} ,
\eqno (2.4)
$$
with associated torsion ${\cal T}_{ab}^{\;\; c}$
$$
2 {\cal D}_{[a}{\cal D}_{b]} f
\equiv {\cal T}_{ab}^{~~c} \; \nabla_c f\, , \,
f\, {\rm a~ zero-form} ,
\eqno (2.5)
$$
$$
{\cal T}_{ab}^{\;\; c}
\equiv T_{ab}^{\;\; c} - 2 P_{\;\;[ab]}^{c} .
\eqno (2.6)
$$
By requiring the annihilation of the
symplectic form $\epsilon_{AB}$ one
gets a restriction on $Q_{aAB}$ above:
$$
{\cal D}_a \epsilon_{AB} = \nabla_a \epsilon_{AB} =0
\;\;\;\; \Rightarrow Q_{aAB} = Q_{a(AB)} \;\; ({\rm trace-free}).
\eqno (2.7)
$$
Concerning the action on space-time indices, and thus $P^c_{\;\; ab}$,
it is known that to control both metricity
condition and torsion it is
necessary to include kinetic terms for them in the corresponding
action [12-13]; otherwise one should impose either
of them and get the other as an equation of motion [12-13].
We follow the latter possibility by imposing
$$
P^c_{\;\; ab} = 0 .
\eqno (2.8)
$$
This amounts to specify that
the torsion of $\nabla$, $T_{ab}^{\;\;\; c}$, is exactly that of
$\cal D$, ${\cal T}_{ab}^{\;\;\; c}$ (cf. (2.6)),
to be determined dynamically. Also, from (2.4),
the action on space-time indices of both $\cal D$ and $\nabla$ is
identified.

Varying $\cal D$ is equivalent to varying $Q$ so
the action (2.2) should be re-expressed in terms of
$\nabla,\; Q_{aMN}$ and $T_{ab}^{\;\;\;\; c}$.
The curvatures, $F_{abMN}$ and $R_{abMN}$ of
${\cal D}$ and $\nabla$ (on unprimed indices),
respectively, are such that [10]
$$
F_{abMN} = R_{abMN} - 2 \nabla_{[a} Q_{b]MN}
+ 2 Q_{[aM}^{~~~~P} \; Q_{b]PN}
+ T_{ab}^{\;\;\; c} \; Q_{cMN} .
\eqno (2.9)
$$
On inserting (2.9) into (2.2), an integration by parts is
necessary to deal with the $\nabla_{[a}Q_{b]MN}$ term.
This gives a total divergence and a term containing
the derivative of products of soldering forms
$$ \eqalignno{
\; &{\int_{M} d^{4}x \left\{-2 \sigma
\; \sigma^{aMA'} \sigma^{bN}_{\;\;\;\; A'}
\nabla_{[a} Q_{b]MN} \right\}}\cr
&= \int_{M} d^{4}x \left\{- 2 \nabla_{a}
\left[\sigma \; \sigma^{[aMA'}
\sigma^{b]N}_{\;\;\;\; A'} \; Q_{bMN} \right]
+ 2 \left[ \nabla_a \left(\sigma \; \sigma^{[aMA'}
\sigma^{b]N}_{\;\;\;\; A'}
\right)\right] Q_{bMN} \right\} \cr
&= - 2 \int_{\partial M} dS_{a}
\; \sigma^{[aMA'} \sigma^{b]N}_{\;\;\;\; A'}
\; Q_{bMN}
- 2 \int_{M} d^{4}x \; T_{am}^{\;\;\; m} \;
\sigma \; \sigma^{[aMA'}
\sigma ^{b]N}_{\;\;\; A'} \; Q_{bMN} .
&(2.10)\cr}
$$
The second term on the second line above drops by virtue of
the metricity condition,
whereas the total divergence turns into a sum of a boundary and
a volume term, the latter containing torsion.

Using the above results in varying (2.2)
w.r.t. $ Q_{gMN}$ yields the equation
$$
\sigma^{[a(MA'} \sigma^{b]N)}_{~~~A'}
\Bigr(2 T_{am}^{~~m} \; \delta_b^{~g}
-  T_{ab}^{~~g} \Bigr)
+ 4 \sigma^{[g(MA'} \sigma^{a]}_{\;\;\; AA'} \; Q_{a}^{\;\; N)A}
+ i \sigma^{g(M}_{~~~A'} \; k^{N)A'} = 0 .
\eqno (2.11)
$$
Here $[a(MA' \; b]N)$ means antisymmetrization in $a,b$ and
symmetrization in $M,N$, and similarly for the other terms.
$k^{AA'}$ and $k^m$ are defined by
$$
k^{AA'} \equiv - i \sqrt{2} \left( \overline{\kappa}^{A'}\kappa^A
- \mu^A \overline{\mu}^{A'}\right) ,
\eqno (2.12a)
$$
$$
k^m \equiv \sigma^m_{\;\; AA'} \; k^{AA'} .
\eqno (2.12b)
$$
One readily solves (2.11) for $Q_{aMN}$ pointing out that
$i\sigma^{g(M}_{\;\;\;\;\;A'} \; k^{N)A'} =
i \sigma^{g(M}_{\;\;\;\;\;A'} \; \epsilon_R^{\;\;N)} k^{RA'} $,
whose r.h.s., in turn, obeys the identity
$$
2 i \sigma^{g(M}_{\;\;\;\;\;\; A'} \; \epsilon_{R}^{\;\; N)}
= \sigma^{p(MB'} \sigma^{qN)}_{\;\;\;\; A'} \; \sigma^m_{\;\;RB'}
\; \epsilon_{pqm}^{\;\;\;\;\;\; g} .
\eqno (2.13)
$$
Note that there is an implicit
antisymmetrization in $p,q$ on the r.h.s.
of this identity owed to the contraction with the volume four-form.
It is possible now to factor out the soldering-form factors in
(2.11). This leads to
$$ \eqalignno{
\sigma^{pRA'} \sigma^{qS}_{\;\;\;\; A'}
&\left\{
\left(2T_{[pm}^{\; \; \; \; \; m}
\; \delta_{q]}^{\;\;\; g} - T_{pq}^{\;\;\; g}
\right) \epsilon_R^{\;\; (M} \epsilon_S^{\;\; N)} \right. \cr
&\left. + 4 \epsilon_{SA} \delta_{[p}^{\; \; \; g}
Q_{q]}^{\; \; \; A(N} \epsilon_R^{\;\; M)}
+ {1\over 2} \epsilon_{pqm}^{\;\;\;\;\;\; g} \; k^{m}
\epsilon_R^{\;\;(M} \epsilon_S^{\;\; N)}
\right\} = 0 .
&(2.14)\cr}
$$
Assuming $\sigma^{aAB'}$ is non-degenerate enables one to set to zero
the factor in braces. Tracing of such a factor over $R,M$ then yields
$$
\epsilon_S^{\;\; N}
\left(2 T_{[pm}^{\; \; \; \; m} \; \delta_{q]}^{\;\;\; g}
-T_{pq}^{\;\;\; g}\right)
- 4 Q_{[qS}^{\; \; \; \; \; N} \; \delta_{p]}^{\;\;\;\; g}
+ {1\over 2} \epsilon_{pqm}^{\;\;\;\;\;\; g}
\; k^{m} \; \epsilon_S^{\;\; N} = 0 .
\eqno (2.15)
$$
Since $Q$ is traceless, taking the traces over $S,N$  and
$q,g$, one finds
$$
T_{am}^{\;\;\; m} =0 ,
\eqno (2.16)
$$
so that torsion takes the value
$$
T_{pq}^{\;\;\; g} =
{1\over 2} \epsilon_{mpq}^{\;\;\;\;\;\; g} \; k^{m} .
\eqno (2.17)
$$
Moreover, on inserting this value of torsion in (2.15) one finds
$$
Q_{aSN} =0 .
\eqno (2.18)
$$
Hence, according to (2.9),
${\cal D}$ is the self-dual part of the connection $\nabla$,
and the corresponding torsion is (2.17) by virtue of
(2.6) and (2.8).

Reproducing (2.1) from (2.2) is easy at this stage.
Recall that the Bianchi symmetry of the curvature of a connection
$\nabla$ having torsion $T$ (see e.g. [10])
$$
R_{[abc]}^{~~~~d} - T_{[ab}^{~~e}
\; T_{c]e}^{~~d} - \nabla_{[a} T_{bc]}^{~~d}
=0 ,
\eqno (2.19)
$$
where antisymmetrization is understood on all three indices $a,b,c$,
can be related to the self-dual Riemann tensor [1]
$$
{ }^{+}R_{abc}^{~~~d} = {1\over 2} \left(
\delta_c^m \delta_n^d - {i\over 2} \epsilon_{c~~~n}^{~dm~} \right)
R_{abm}^{~~~~n}
= R_{abA}^{~~~~B} \; \sigma_c^{~AM'} \sigma_{~BM'}^{d} .
\eqno (2.20)
$$
Such a relation is as follows. The self-dual scalar curvature providing the
total pure-gravity contribution to the self-dual action can be written as
$$
{ }^{+}R \equiv  \delta_d^{\;\; b} g^{ac}\;
{ }^{+}R_{abc}^{\; \; \; \; \; d}
= {1\over 2} R - {i\over 4} \epsilon^{abm}_{\;\;\;\;\;\;n} \;
R_{abm}^{\;\;\;\;\;\;n} .
\eqno (2.21)
$$
The second term of the last equality can be obtained by means of
the Bianchi symmetry (2.19) and of the torsion (2.17) as
$$
\epsilon_d^{\;\;abc} \; R_{abc}^{\; \; \; \; \; d}
= \epsilon_d^{\;\;abc} \; \nabla_{[a} T_{bc]}^{\; \; \; d}
= 3 \nabla_{a} k^{a} ,
\eqno (2.22)
$$
since the term quadratic in torsion drops out in view of the form
of the torsion (2.17). Finally, one obtains
$$
{ }^{+}R = {1\over 2} R + {3i\over 4} \nabla_{a} k^{a} .
\eqno (2.23)
$$
Correspondingly, the terms containing derivatives of the
fermionic fields in the self-dual
action (2.2) can be re-written in terms of $\nabla$ and
$k^m$ as follows:
$$ \eqalignno{
-\sqrt{2} \; \sigma \; \sigma^{a}_{~AA'} \left[
\overline{\kappa}^{A'}
\left({\cal  D}_a \kappa^A \right)
-\left({\cal  D}_a \mu^A\right)
\overline{\mu}^{A'}\right]
&= - {\sigma \over \sqrt{2}} \sigma^{a}_{~AA'} \left[
\overline{\kappa}^{A'} \left(\nabla_a \kappa^A\right)
- \left(\nabla_a \mu^A\right)
\overline{\mu}^{A'}\right]\cr
& + {\sigma \over \sqrt{2}} \sigma^{a}_{~AA'} \left[
\left(\nabla_a \overline{\kappa}^{A'} \right)\kappa^{A}
- \mu^A \left(\nabla_a \overline{\mu}^{A'}\right)
\right] \cr
& - {i \over 2} \sigma \nabla_{a} k^{a} .
&(2.24)\cr}
$$
In the light of (2.23)-(2.24) we have
shown that the ECSK-Dirac action (2.1) and
the self-dual action (2.2) are equivalent modulo
total divergences and the equation of motion for $\cal D$ (i.e.
$\cal D$ is the self-dual part of $\nabla$). Note that
the mass terms in the actions differ by a factor of $2$.
Because of the non-vanishing torsion of $\nabla$ these divergences give,
apart from the boundary terms, volume terms involving the trace of
the torsion. However, for the Einstein-Dirac system, torsion is traceless
(see (2.16)) and hence we get, indeed, a complete
dynamical equivalence. Explicitly,
$$
S_{SD}\Bigr[{ }^{+}\omega(\sigma,T),\sigma,\kappa,\mu \Bigr] =
{1\over 2} S_{ECSKD}\Bigr[\omega(\sigma,T),\sigma,\kappa,\mu \Bigr]
+{i \over 4}\int_{\partial M}dS^{a} k_{a} ,
\eqno (2.25)
$$
${ }^{+}\omega(\sigma,T)$ being the self-dual part
of the connection $\omega(\sigma,T)$; the arguments, soldering form
and torsion, indicating their equations of motion
(cf. (2.4), (2.8) and (2.18)), have been used. For
real general relativity, it is then evident that,
although $S_{SD}$ is not real, its imaginary
part is a boundary term. This is a non-trivial
generalization to spin-${1\over 2}$ fields
coupled to gravity of the results obtained in [2] for pure gravity
(cf. (1.3)).
\vskip 0.3cm
\leftline {\bf 3. - Field equations.}
\vskip 0.3cm
The form of the action (2.2) makes it easy
to get the equations of motion for the remaining fields.
By varying with respect to $\mu^A,{\overline{\mu}^{A'}},
{\kappa^A},{\overline{\kappa}^{A'}}$
and using $\widetilde{\sigma}^a_{\;\;AA'}
\equiv \sigma \; \sigma^a_{\;\;AA'}$,
the equations of motion for the Dirac field are
$$
\widetilde{\sigma}^a_{\;\; AA'} {\cal D}_a\kappa^A
={im \over \sqrt 2} \sigma \; \overline{\mu}_{A'}
\; \; \; \; \; \; \; \;
{\cal D}_a \left(\widetilde{\sigma}^a_{\;\;AA'}
\overline{\kappa}^{A'} \right)
={im \over \sqrt{2}} \sigma \; \mu_{A} ,
\eqno (3.1a)
$$
$$
{\cal D}_a \left(\widetilde{\sigma}^a_{\;\;AA'}\overline{\mu}^{A'} \right)
={im \over \sqrt{2}} \sigma \; \kappa_{A}
\; \; \; \; \; \; \; \;
\widetilde{\sigma}^a_{AA'} {\cal D}_a\mu^A
={im \over \sqrt 2} \sigma \; \overline{\kappa}_{A'} .
\eqno (3.1b)
$$
Note that $\cal D$ does not act on primed indices
(hence its compatibility with the soldering form is undefined)
but one needs to know
its action on space-time indices, whereas in the pure-gravity case
it is independent of its extension to act on space-time indices
because of the torsion-free condition [1]. Here, however,
it develops a non-vanishing torsion.
This problem is automatically solved by our request that the
connection $\nabla$ should coincide with
$\cal D$ when acting on space-time
indices and hence should have identical torsion (see (2.4), (2.8)).

To compare with the standard Dirac equations of motion we simply
replace $\cal D$ with $\nabla$. This leads to
$$
\sigma^a_{\;\; AA'} \nabla_a\kappa^A
={im \over \sqrt{2}} \overline{\mu}_{A'}
\; \; \; \; \; \; \; \;
\sigma^a_{\;\; AA'}\nabla_a \overline{\kappa}^{A'}
={im \over \sqrt{2}} \mu_{A} ,
\eqno (3.2a)
$$
$$
\sigma^a_{\;\; AA'}\nabla_a \overline{\mu}^{A'}
={im \over \sqrt{2}} \kappa_A
\; \; \; \; \; \; \; \;
\sigma^a_{\;\; AA'} \nabla_a\mu^A
= {im \over \sqrt{2}} \overline{\kappa}_{A'} .
\eqno (3.2b)
$$
With our conventions $\sigma^a_{\;AA'}$ are taken to be antihermitian.
Although equations (3.2a)-(3.2b) resemble ordinary
Dirac equations in curved space-time, one
should bear in mind $\nabla$ is {\it not} torsion-free.
These are the equations for a Dirac field minimally coupled
to gravity with torsion (see e.g. [9]).
The rest of the field equations requires varying w.r.t. $\cal D$ and
$\sigma^a_{\;\;AA'}$. As usual in the Palatini formalism, the former
variation yields the value of the torsion (section 2) whereas the latter
leads to the Einstein (-Cartan) equations with source the spin-${1\over 2}$
field [7,9], i.e.
$$
G_{ab}={1\over\sqrt{2}} \sigma_{bAA'}  \left[
\overline{\kappa}^{A'} \left(\nabla_a \kappa^A\right)
-\left(\nabla_a\overline{\kappa}^{A'}\right)\kappa^A
+\mu^A \left(\nabla_a\overline{\mu}^{A'}\right)
-\left(\nabla_a\mu^A\right) \overline{\mu}^{A'} \right] \;.
\eqno (3.3)
$$
It is now possible to make contact with the results of [5].
The authors of this reference
found a cubic term in fermionic fields
in their Dirac equation and stressed it has its origin in the
kind of theory they started with, i.e. the torsion in the
first-order theory (2.2) by analogy with the ECSKD action (2.1).
This is explicitly shown below by splitting out the torsion
contribution from the connection $\nabla$ introduced above.
Let ${\widetilde \nabla}$ be the unique torsion-free connection
compatible with the metric $g_{ab}=\sigma_{aAA'}\sigma_b^{AA'}$.
Hence, there exists a tensor
$Q_{ab}^{\;\;\;\;c}$, and its spinor version $\Theta_{aBC}$,
$\overline{\Theta}_{aB'C'}$, relating
${\widetilde \nabla}$ and $\nabla$ through [10]
$$
\left({\widetilde \nabla}_a - \nabla_a\right) v^b
=Q_{ac}^{\;\;\;\;b} \; v^{c} ,
\eqno (3.4)
$$
$$
\left({\widetilde \nabla}_a - \nabla_a\right) \kappa^A
=\Theta_{a\;\;B}^{\;\;A} \; \kappa^{B} ,
\eqno (3.5)
$$
$$
\left({\widetilde \nabla}_a - \nabla_a\right) \lambda^{A'}
=\overline{\Theta}_{a\;\;B'}^{\;\;A'} \; \lambda^{B'} ,
\eqno (3.6)
$$
where the spinor decomposition
$$
Q_{ab}^{\;\;\;\;c} = \left[\Theta_{a\;\;B}^{\;\;C}
\; \epsilon_{B'}^{\;\;C'}
+ \overline{\Theta}_{a\;\;B'}^{\;\;C'}
\; \epsilon_{B}^{\;\;C}\right]
\; \sigma_{b}^{BB'} \; \sigma_{\; CC'}^{c} ,
\eqno (3.7)
$$
$$
\Theta_{aBC} = {1\over 2} \sigma^b_{BB'} \; \sigma^{c\;\;B'}_{\;\;C}
Q_{abc} ,
\eqno (3.8)
$$
is implied. With our notation, $\Theta_{aBD}$ corresponds to
$C_{aBD}$ appearing in [5]. Furthermore
$$
T_{ab}^{\;\;\;\;c} = 2 Q_{[ab]}^{\;\;\;\;c} ,
\eqno (3.9)
$$
$$
Q_{abc} = T_{a[bc]} - {1\over 2} T_{bca} ,
\eqno (3.10)
$$
the first of which states that ${\widetilde \nabla}$ is
torsion-free and the second is
a result of the metricity condition [10]. The torsion
information is hence contained in
$\Theta_{aBC}$. In the case of Dirac
fields one gets, inserting (2.17) into the above relations,
$$
\Theta_{aBC} = {i \over 4}
k_{(BA'} \; \sigma_{aC)}^{\;\;\;\;\;\; A'} .
\eqno (3.11)
$$
Hereby the modification to the Dirac equation found in [5]
is explicitly determined, its origin being the non-vanishing torsion;
by virtue of (3.5) and (3.11), the
first of the Dirac equations (3.2a) takes the form
$$
\sigma^a_{\;\;AA'} \nabla_{a} \kappa^{A}
= \sigma^{a}_{\;\;AA'}
\left({\widetilde \nabla}_{a}
- {3i \over 8} k_{a}\right) \kappa^{A} \;
= {im \over \sqrt{2}} {\overline \mu}_{A'} .
\eqno (3.12)
$$
This result extends to the primed-indices spinor equations
(3.2) through $\overline{\Theta}_{aB'C'}$, and, similarly,
to the Einstein equations (3.3). Such a modification
was discussed previously in a $U_4$-theory [9].
The reduced action of [5] is thereby obtained. In
particular, the four-Fermi interaction term, $\sigma k_mk^m$, is
brought into the action; in other words, using the space-time-indices
version of the identity (2.9)
(cf. [10]) for the curvatures of
$\nabla$ and ${\widetilde \nabla}$,
one gets the following relation between the corresponding scalars:
$
R={\widetilde R}-{3 \over 8}k_{m}k^{m}
$.
\vskip 0.3cm
\leftline {\bf 4. - Concluding remarks.}
\vskip 0.3cm
Our analysis proves
the equivalence between the self-dual and the ECSK forms of
the action coupling Dirac fields to gravity,
by introducing a connection with
non-vanishing torsion. The key steps are the use of the
Bianchi symmetry of the curvature of such a connection,
here including torsion, and the result that the torsion
for this system is
totally antisymmetric. Thus, the actions differ by total divergences.
They lead to boundary terms only, since the volume terms they
involve are proportional to the trace of the torsion, and hence
vanish. This can be considered the explicit version
of an observation first made by Dolan [14].
He studied the canonical transformation, in pure gravity, from
tetrad and connection variables
to Ashtekar's variables. According to [14],
the generating function, when torsion is present,
has the same structure, whenever torsion is totally
antisymmetric; this is the case for ED theory and supergravity.
On the other hand,
Jacobson [4] used another approach to prove
the above equivalence of the actions. The boundary terms he finds,
can thus be traced back to the Bianchi
identity by using the present results
(see the second term in (2.25)).

Moreover, we have shown explicitly that the extra term
entering the Dirac equations obtained in
[1,5] from the self-dual action
is a torsion term. By splitting the self-dual connection
into its torsion-free and torsion parts, the standard
four-Fermi interaction in the action is obtained
[4-5]. These results completely agree with [9],
where the authors investigated the
four-Fermi interaction using a certain anholonomic
basis and the corresponding connection.

We are currently investigating the holomorphic version of the
ECSKD theory, motivated by the complex space-time program of
Penrose [15]. The corresponding theory appears to be more rich
than the usual ECSKD theory studied in canonical gravity and in
our paper, and it deserves further study to shed new light on
complex general relativity and quantum gravity.
\vskip 1cm
\centerline {* * *}
\vskip 1cm
\noindent
The authors are indebted to D. W. Sciama for
drawing their attention to the problem of
torsion. H. A. Morales-T\'ecotl
is grateful to G. F. R. Ellis, T. Jacobson and J. D. Romano
for useful hints and several comments,
and E. T. Newman and A. Ashtekar for their kind hospitality during a
visit to Pittsburgh and Syracuse Universities, where part of this work was
performed. He was supported by the Italian M.U.R.S.T and by CONACyT
Grants 55751 and 3544-E9311. G. Esposito is indebted to the Italian
M.U.R.S.T. for travel funds.
\vskip 1cm
\leftline {REFERENCES}
\vskip 1cm

\item {[1]}
A. ASHTEKAR: {\it Lectures on Non-Perturbative
Canonical Gravity} (World Scientific, Singapore, 1991).
\item {[2]}
T. JACOBSON and L. SMOLIN: {\it Phys. Lett. B}, {\bf 196}, 39 (1987);
T. JACOBSON and L. SMOLIN: {\it Class. Quantum Grav.}, {\bf 5},
583 (1988).
\item {[3]}
C. ROVELLI: {\it Class. Quantum Grav.}, {\bf 8}, 297 (1991);
C. ROVELLI: {\it Class. Quantum Grav.}, {\bf 8}, 317 (1991);
C. ROVELLI: {\it Phys. Rev. D}, {\bf 43}, 442 (1991).
\item {[4]}
T. JACOBSON: {\it Class. Quantum Grav.}, {\bf 5}, L143 (1988).
\item {[5]}
A. ASHTEKAR, J. D. ROMANO and R. TATE: {\it Phys. Rev. D},
{\bf 40}, 2572 (1989).
\item {[6]}
T. W. B. KIBBLE: {\it J. Math. Phys.} (N.Y.), {\bf 2}, 212 (1961);
D. W. SCIAMA: {\it On the analogy between charge and spin in
general relativity}, in {\it Recent Developments in
General Relativity} (Pergamon Press, Oxford, 1962).
\item {[7]}
F. W. HEHL, P. von der HEYDE and  G. D. KERLICK:
{\it Rev. Mod. Phys.}, {\bf 48}, 393 (1976).
\item {[8]}
J. E. NELSON and C. TEITELBOIM: {\it Ann. Phys.}
(N.Y.), {\bf 116}, 86 (1977).
\item {[9]}
F. W. HEHL and B. K. DATTA: {\it J. Math. Phys.}
(N.Y.), {\bf 12}, 1334 (1971).
\item {[10]}
R. PENROSE and W. RINDLER: {\it Spinors and Space-Time I}
(Cambridge University Press, Cambridge, 1984).
\item {[11]}
G. ESPOSITO, H. A. MORALES-T\'ECOTL and G. POLLIFRONE:
{\it Found. Phys. Lett.}, {\bf 7}, 303 (1994).
\item {[12]}
P. A. M. DIRAC: {\it Interacting gravitational and spinor fields},
in {\it Recent Developments in General Relativity}
(Pergamon Press, Oxford, 1962).
\item {[13]}
R. FLOREANINI and R. PERCACCI:
{\it Class. Quantum Grav.}, {\bf 7}, 1805 (1990).
\item {[14]}
B. P. DOLAN: {\it Phys. Lett. B}, {\bf 233}, 89 (1989).
\item {[15]}
R. PENROSE and W. RINDLER: {\it Spinors and Space-Time II}
(Cambridge University Press, Cambridge, 1986).

\bye